\documentclass[a4paper]{jpconf}

\usepackage[dvips]{epsfig}
\usepackage{amsfonts}
\usepackage{amssymb}
\usepackage{amscd}
\usepackage{pstricks}
\usepackage{amstext}
\input xy
\xyoption{all}

\newtheorem{example}{Example(s)}[section]

\newtheorem{definition}[example]{Definition}

\newtheorem{remark}[example]{Remark}

\newtheorem{exemple}[example]{Example}
\newtheorem{conjecture}[example]{Conjecture}
\newtheorem{resultat}[example]{Result}

\def\hs{\hbox to 3mm{}}
\def\hhs{\hbox to 5cm{}}

\def\ms{\smallskip}

\def\adots{\mathinner{\mkern2mu\raise1pt\hbox{.}
\mkern3mu\raise4pt\hbox{.}\mkern1mu\raise7pt\hbox{.}}}

\def\up#1{\raise 1ex\hbox{\footnotesize#1}}

\def\N{{\mathbb N}}
\def\C{{\mathbb C}}

\begin{document}

\title{Approximate substitutions and the normal ordering
problem}

\author{H Cheballah$^{\dag}$,
G H E Duchamp$^{\dag}$ and K A Penson$^{\diamondsuit}$ }
\address{
$^\dag$ Universit\'e Paris 13\\ Laboratoire d'Informatique Paris
Nord, CNRS UMR 7030\\99 Av. J-B. Cl\'ement, F 93430 Villetaneuse,
France\vspace{2mm} }
\address
{$^\diamondsuit$ Laboratoire de Physique Th\'eorique de la Mati\`ere Condens\'ee\\
Universit\'e Pierre et Marie Curie, CNRS UMR 7600\\
Tour 24 - 2e \'et., 4 pl. Jussieu, F 75252 Paris Cedex 05, France}

\eads{\linebreak\mailto{hayat.cheballah@lipn-univ.paris13.fr},
\mailto{ghed@lipn-univ.paris13.fr},
\mailto{penson@lptl.jussieu.fr} \linebreak}

\begin{abstract}
In this paper, we show that the infinite generalised Stirling
matrices associated with boson strings with one annihilation
operator are projective limits of approximate substitutions, the
latter being characterised by a finite set of algebraic equations.
\end{abstract}

\section{Introduction}
The series of papers \cite{Blasiak,GBNOP,CPNOMFG} had two sequels.
First one, algebraic, was the construction of a Hopf algebra  of
Feynman-Bender diagrams \cite{GOF10,GOF11}  arising from the
product formula applied to two exponentials. Second one was the
construction and description of one parameter groups of infinite
matrices \cite{Gentle_intro,one_param_group} and their link with
the combinatorics of so called \textit{Sheffer polynomials}. The
object of this paper is to continue the investigation of those
\textit{one-parameter groups} by highlighting the structure of the
group of substitutions.\\ It is shown here how we can see this
group as a projective limit  of what will be called
\textit{approximate substitution groups}.

First, we consider Boson creation and annihilation operators with
the commutation relation
\begin{eqnarray}\label{[a,a+]=1}
  [a,a^\dag] &=& 1
\end{eqnarray}
Recall that \cite{Blasiak}
\begin{itemize}
    \item the annihilation operator $a$, in the second quantization,
    represents an operator which changes each state $|N\rangle$ of the Fock space
    (containing $N\geq 1$) particles
to another containing $N-1$ particles. One has here
\begin{eqnarray}
    a|N\rangle &=& \sqrt{N}|N-1\rangle
\end{eqnarray}

    \item the hermitian conjugate of the annihilation operator is the creation
operator $a^\dag$ which changes each state $|N\rangle$ of the Fock
space containing $N$ particles to another containing $N+1$
particles. One has then
\begin{eqnarray}
  a^\dag|N\rangle &=& \sqrt{N+1}|N+1\rangle
\end{eqnarray}

\end{itemize}

Starting from the vacuum $|0\rangle$, we can reach all the
normalized states in the Fock space. There are indeed given by:
\begin{eqnarray}
  |N\rangle &=&
  \frac{(a^\dag)^{N}}{\sqrt{N!}}|0\rangle
\end{eqnarray}

\section{Normally ordered form}
The noncommutativity of annihilation and creation operators may
cause problems in defining an operator function in quantum
mechanics. To solve these problems, we have to find some suitable
form which allows computing, reduction and equality test. One of
the simplest and widely used procedure is the finding of the
\textit{normally ordered form} of the boson operators in which all
$a^\dag$ stand to the left of all the factors $a$
\cite{Blasiak}.\\There are two well known procedures defined on
the boson expressions: namely $\mathcal N$, the \textit{normal
ordering}
 and  $::$ , the \textit{double dot operation}, \cite{Gentle_intro}.
\subsection{Normal ordering}
By  normal ordering of a general expression $F(a^\dag,a)$, we mean
 $\mathcal{N}[F(a^\dag,a)]$ which is obtained by moving all the
 annihilation operators $a$ to the right, using the commutation
 relation (\ref{[a,a+]=1}).
 \begin{exemple} Let  $w\in\{a,a^\dag\}^*$ a word given by
$w=aa^\dag aaa^\dag a$. The normal ordering of $w$ is
\begin{eqnarray*}
   aa^\dag aaa^\dag a &=&  (1+a^\dag a)a(1+a^\dag a)a=a^2+a^\dag a^3+aa^\dag a^2+a^\dag a^2 a^\dag a^2\\
   &=& a^2+a^\dag a^3+(1+a^\dag a)a^2+a^\dag a(1+a^\dag a)a^2\\
   &=& a^2+a^\dag a^3+a^2+a^\dag a^3+a^\dag a^3+a^\dag aa^\dag
   a^3\\
   &=& 2 a^2+3a^\dag a^3+a^\dag(1+a^\dag a)a^3\\
   &=& 2a^2+3a^\dag a^3+(a^\dag)^2a^4
\end{eqnarray*}
 \end{exemple}
 \begin{remark}
Note that all coefficients of the normal ordering of a word (more
precisely the coefficients of the decomposition of a word
 on the basis $\Big\{(a^\dag)^j a^l\Big\}$)
are positive integers. This suggests that these integers count
combinatorial objects \cite{url}.
 \end{remark}
\subsection{Double dot operation}
The double dot operation $:F(a^\dag,a):$ is a similar procedure
using another commutation relation \textit{i.e.} $[a,a^\dag]=0$
instead of $[a,a^\dag]=1$, \textit{i.e} moving all annihilation
operators $a$ to the right as if they were commuting with
the creation operators $a^\dag$.\\
\begin{remark}The double dot operation $: :$ is a linear operator
which can be directly defined, for a word $w\in\{a,a^\dag\}^*$,
by:
$$:w:=a^{\dag(|w|_{a^\dag})}a^{(|w|_{a})}$$
where $|w|_x$ stands for the number of occurences of a symbol $x$
in the word $w$.
\end{remark}
\begin{exemple} We take the same word as above:
 $w=aa^\dag aaa^\dag a$.\\
The double dot operation gives
\begin{eqnarray*}
:aa^\dag aaa^\dag a: &=& a^\dag a^\dag aaaa
\end{eqnarray*}
\end{exemple}

\section{Combinatorics of the normal ordering}
The Bell and Stirling numbers have a purely combinatorial origin
\cite{comtet}, but in this communication, we will consider them as
coefficients of the normal ordering problem~ \cite{Blasiak}.\\
The general word $w\in\{a,a^\dag\}^*$ with letters in
$\{a,a^\dag\}$, \textit{i.e.} Boson string, can be described by
two sequences of non negative integers $\textbf{r}=(r_1,r_2,\cdots
,r_M)$ and $\textbf{s}=(s_1,s_2,\cdots ,s_M)$, so that we define
\begin{eqnarray}\label{H_rs_def}
   w_{\textbf{r},\textbf{s}} &=&
   (a^\dag)^{r_1}a^{s_1}(a^\dag)^{r_2}a^{s_2}\cdots (a^\dag)^{r_M}a^{s_M}
\end{eqnarray}
and $\displaystyle{d=\sum_{m=1}^n(r_m-s_m),\quad n=1,\cdots , M}$
represents the excess (\textit{i. e.} the difference between the
number of creations and the number of annihilations). Then, the
normally ordered form of $w_{\textbf{r},\textbf{s}}^n$ is
given by\vspace{0.5mm}\\

$\mathcal N(w_{\textbf{r},\textbf{s}}^n)=$$\left\{%
\begin{array}{ll}
     \displaystyle{(a^\dag)^{nd}\sum_{k=0}^{\infty}
S_{\textbf{r},\textbf{s}}(n,k)(a^\dag)^ka^k,} & \hbox{if $d>0$;} \\
     \displaystyle{\Big(\sum_{k=0}^{\infty}
S_{\textbf{r},\textbf{s}}(n,k)(a^\dag)^ka^k\Big) (a^\dag)^{n|d|},} & \hbox{otherwise,} \\
\end{array}%
\right.$ 

\bigskip
where quantities $S_{\textbf{r},\textbf{s}}(n,k)$ are
generalizations of standard Stirling
numbers \cite{GBNOP,CPNOMFG}.\\
The generalized Bell polynomials $B_{\textbf{r},\textbf{s}}(n,x)$
and the generalized Bell numbers $B_{\textbf{r},\textbf{s}}(n)$
are defined respectively by
\begin{eqnarray}
  B_{\textbf{r},\textbf{s}}(n,x) &=& \sum_{k=0}^\infty S_{\textbf{r},\textbf{s}}(n,k) x^k=
  \sum_{k=0}^{nr} S_{\textbf{r},\textbf{s}}(n,k) x^k \\
  B_{\textbf{r},\textbf{s}}(n,1) &=& B_{\textbf{r},\textbf{s}}(n)=
  \sum_{k=0}^\infty S_{\textbf{r},\textbf{s}}(n,k)=\sum_{k=0}^{nr} S_{\textbf{r},\textbf{s}}(n,k)
\end{eqnarray}
\begin{remark} Notice that\vspace{0.5cm}

$S_{\textbf{r},\textbf{s}}(n,k)=$$\left\{%
\begin{array}{ll}
     1, & \hbox{for $k=nr$;} \\
     0, & \hbox{for $k>nr$.} \\
\end{array}%
\right.$ \vspace{0.5cm}
\end{remark}
\begin{exemple} Using a computer algebra program, we get the
following matrices.\\

\begin{description}
    \item  [$\bullet\quad w=a^\dag a$]$ $, we get the usual matrix of
\textit{Stirling numbers of second kind} $S(n,k) \cite{comtet}$\\

    $$
\left(%
\begin{array}{cccccccc}
  1 & 0 & 0 & 0 & 0 & 0 & 0 & \cdots  \\
  0 & 1 & 0 & 0 & 0 & 0 & 0 & \cdots \\
  0 & 1 & 1 & 0 & 0 & 0 & 0 & \cdots \\
  0 & 1 & 3 & 1 & 0 & 0 & 0 & \cdots \\
  0 & 1 & 7 & 6 & 1 & 0 & 0 & \cdots \\
  0 & 1 & 15 & 25 & 10 & 1 & 0 & \cdots \\
  0 & 1 & 31 & 90 & 65 & 15 & 1 & \cdots \\
 \vdots & \vdots & \vdots & \vdots & \vdots & \vdots & &\ddots \\
\end{array}%
\right)
$$

    \item[$\bullet\quad w=a^\dag aa^\dag$], we have\\
$$
\left(%
\begin{array}{cccccccc}
  1 & 0 & 0 & 0 & 0 & 0 & 0 & \cdots  \\
  1 & 1 & 0 & 0 & 0 & 0 & 0 & \cdots \\
  2 & 4 & 1 & 0 & 0 & 0 & 0 & \cdots \\
  6 & 18 & 9 & 1 & 0 & 0 & 0 & \cdots \\
  24 & 96 & 72 & 16 & 1 & 0 & 0 & \cdots \\
  120 & 600 & 600 & 200 & 25 & 1 & 0 & \cdots \\
  720 & 4320 & 5400 & 2400 & 450 & 36 & 1 & \cdots \\
 \vdots & \vdots & \vdots & \vdots & \vdots & \vdots & &\ddots \\
\end{array}%
\right)
$$

\vspace{1.0cm}

    \item [$\bullet\quad w=a^\dag aaa^\dag a^\dag$], one gets\\
    $$
\left(%
\begin{array}{cccccccccc}
  1 & 0 & 0 & 0 & 0 & 0 & 0 & 0 & 0 & \cdots  \\
  2 & 4 & 1 & 0 & 0 & 0 & 0 & 0 & 0 &  \cdots \\
  12 & 60 & 54 & 14 & 1 & 0 & 0 & 0 & 0 &  \cdots \\
  144 & 1296 & 2232 & 1296 & 306 & 30 & 1 & 0 & 0 &  \cdots \\
  2880 & 40320 & 109440 & 105120 & 45000 & 9504 & 1016 & 52 & 1 &  \cdots \\
 \vdots & \vdots & \vdots & \vdots & \vdots & \vdots & \vdots & \vdots & &\ddots \\
\end{array}%
\right)
$$
\end{description}
\end{exemple}

\vspace{0.5cm}

\begin{remark}
In each case, the matrix $(S_{\textbf{r},\textbf{s}}[n,k])_{n\geq
0, k\geq 0}$ is of a staircase form and the dimension of the step
is the number of $a$'s in the word $w$. Thus, all the matrices are
row-finite and are unitriangular iff the number of its
annihilation operators is exactly one. Moreover, the first column
is $(1,0,\cdots ,0,\cdots)$ iff $w$ ends with $a$ (this means that
$\mathcal N(w^n)$ has no constant term for all $n>0$) \cite{Gentle_intro,one_param_group}.\\

A word $w$ with only one annihilation operator can be written in
the unique form $w=(a^\dag)^{r-p}a (a^\dag)^p$. Then:
\begin{itemize}
    \item if $p=0$, $S_w$ is the matrix of a \textit{unipotent
    substitution}.
    \item if $p>0$, $S_w$ is the matrix of a \textit{unipotent substitution with
    prefunction} \cite{one_param_group}.
\end{itemize}
\end{remark}

\section{Approximate substitutions}
We define here the space of transformation matrices and its
topology, and then we concentrate on the \textit{Riordan subgroup}
\cite{matrice} (\textit{i.e} transformations which are
substitutions
with prefactor functions) \cite{Gentle_intro,one_param_group}.\\

Let $\C^\N$ be the vector space of all complex sequences, endowed
with the Frechet product topology \cite{bourbaki2,rudin}. The
algebra $\mathcal L(\C^\N)$ of all continuous operators
$\C^{\N}\longrightarrow \C^{\N}$ is the space of row-finite
matrices with complex coefficients (a subspace of $\C^{\N\times
\N}$). Let $M$ be a matrix of this space. For a sequence
$A=(a_n)_{n\geq 0}$, the transformed sequence $B=MA$ is given by
$B=(b_n)_{n\geq 0}$ with:
\begin{eqnarray}\label{suite-transformée-bn}
  b_n &=& \sum_{k\geq 0} M(n,k)a_k.
\end{eqnarray}
We will also associate to $A$ (see paragraph
(\ref{approxim_subs})) its exponential generating function
\begin{eqnarray}\label{suite_transform_serie_univariée}
   \sum_{n\geq 0}a_n\frac{z^n}{n!}.
\end{eqnarray}
\subsection{Substitutions with prefunctions}
We now examine an important class of transformations: the
substitutions with prefunctions.

We consider, for a generating function $f$, the transformation
\begin{eqnarray}\label{transformation_prefonction}
  \mathcal T_{g,\phi}[f](x) &=& g(x)f\Big(\phi(x)\Big)
\end{eqnarray}
where $g(x)=1+\sum_{n=1}^\infty g_n x^n$ and
$\phi(x)=x+\sum_{n=2}^\infty \phi_n x^n$ are arbitrary formal
power series. The mapping $\mathcal T_{g,\phi}$ is a linear
application \cite{one_param_group}, the matrix of this
transformation $M_{g,\phi}$ is given by the transforms of the
monomials $\frac{x^k}{k!}$ hence
\begin{eqnarray}\label{transformation_matrice}
  \sum_{n\geq 0}M_{g,\phi}(n,k)\frac{x^n}{n!} =
  \mathcal T_{g,\phi}\Big[\frac{x^k}{k!}\Big]=g(x)\frac{\phi(x)^k}{k!}
\end{eqnarray}
\begin{remark}
If $f(x)=e^{yx}$, Eq. (\ref{transformation_matrice}) comes down to
the \textit{Sheffer condition} on the matrix of $\mathcal T$. It
amounts to the statement that \cite{Blasiak}:
\begin{eqnarray}
  \sum_{n,k\geq 0}T(n,k)\frac{x^n}{n!}y^k &=& g(x)e^{y\phi (x)}
\end{eqnarray}
where $T\in\mathcal L(\C^\N)$ is a matrix with non-zero two first
columns.
\end{remark}
\subsection{Approximate substitutions}\label{approxim_subs}

In this section, we define approximate substitutions matrices and
give a way to determine whether an unipotent (lower triangular
with all diagonal elements equal to $1$) matrix is a matrix of an
approximate substitution.
\begin{definition}
Let $M\in\C^{[0\cdots n]\times [0\cdots n]}$ be a unipotent
matrix, $\displaystyle{M=M[i,k]_{0\leq i, k \leq
n}=({a_{ik}})_{0\leq i, k \leq n}}$; is called matrix of
approximate substitution if it satisfies the following condition:
\begin{eqnarray}
  c_k &=& \Big[c_0\frac{\Big(\frac{c_1}{c_0}\Big)^k}{k!}\Big]_n,\quad\quad\mbox{for all}
  \quad\quad 0\leq k\leq n\label{condition de substitution}
\end{eqnarray}
where
$$c_k=\sum_{i=0}^n M[i,k]\frac{x^i}{i!}$$

$$M=\left(%
\begin{array}{ccccccc}
  1 & 0 & 0 & \cdots & \cdots & \cdots & \cdots \\
  a_{1,0} & 1 & 0 & \cdots & \cdots & \cdots & \cdots \\
  a_{2,0} & a_{2,1} & 1 & 0 & \cdots & \cdots & \cdots \\
  a_{3,0} & a_{3,1} & \cdots & 1 & 0 & \cdots & \cdots \\
  a_{4,0} & a_{4,1} & \cdots & a_{4,k} & \cdots & \cdots & \cdots \\
  a_{5,0} &  a_{5,1} &  \cdots & a_{5,k} & \cdots & 1 & \cdots  \\
  \vdots & \vdots & \vdots & \vdots & \vdots & \vdots & \ddots \\
\end{array}%
\right)$$ \vspace{0.3cm}
\hspace{5.7cm}$c_0$\hspace{0.6cm}$c_1$\hspace{1.5cm}$c_k$\vspace{0.5cm}

\vspace{0.3cm}

 Thus $c_k$ represents the exponential
generating series (here a polynomial) of the $k^{th}$ column
(hence $c_0$, $c_1$ are respectively the exponential generating
series of the $1^{st}$ and the $2^{nd}$ column) and
$\displaystyle{\Big[\quad\Big]_n}$ is the truncation, at order
$n$, of a series.
\end{definition}

We consider now the set of matrices with complex coefficients
noted by $\C^{\N\times\N}$ \cite{bourbaki} and let $\C^{[0\cdots
n]\times [0\cdots n]}$ be the set of all matrices of size
$(n+1)\times (n+1)$. Let also $r_n$ be the truncation of the
matrices taking the upper left principal submatrix of dimension
$(n+1)$, hence $\displaystyle{r_n(M)=\Big(M[i,k]\Big)_{0\leq i,
k\leq n}}$. Thus, we get a linear mapping

\vspace{0.5cm}
$$r_n:\C^{\N\times\N}\longrightarrow \C^{[0\cdots n]\times [0\cdots
n]}$$

It is clear that $r_n$ is not a morphism for the (partially
defined) multiplication (\textit{i.e.} $\displaystyle{r_n(AB)\neq
r_n(A)r_n(B)}$ in general).\\We consider now $\mathcal L \mathcal
T(\N,\C)$ the algebra of lower triangular matrices and $\mathcal
L\mathcal T([0\cdots n],\C)$ the matrices of size $([0\cdots
n]\times [0\cdots n])$ obtained by the truncation $\tau_n$. Then
$$\tau_n:\mathcal L\mathcal
T(\N ,\C)\longrightarrow \mathcal L\mathcal T([0\cdots n],\C)$$
and, this time, $\tau_n$ preserves multiplication ($\tau_n$ is a
morphism). One has the diagram
$$
\xymatrix@C=50pt{
\C^{\N\times\N}\ar[r]^{r_n}& \C^{[0\cdots n]\times [0\cdots n]} \\
 \mathcal L\mathcal
T(\N ,\C)\ar[r]^{\tau_n}\ar[u]^{\mathcal J_1} & \mathcal L\mathcal
T([0\cdots n],\C)\ar[u]^{\mathcal J_2} }
$$
where $\mathcal J_1$ and $\mathcal J_2$ are two canonical
injections.
\begin{remark}
We can write
\begin{eqnarray}
  \mathcal L\mathcal
T(\N,\C) &\displaystyle{=\lim_{\longleftarrow}}& (\mathcal
L\mathcal T([0\cdots n],\C))
\end{eqnarray}
which means that $\mathcal L\mathcal T([0\cdots n],\C)$ is the
projective limit of $ \mathcal L\mathcal T(\N,\C)$
\end{remark}
\subsection{Random generation}
Our motivation, in this section, consists in approximating the
matrices of infinite substitutions by finite matrices of
(approximate) substitutions.  We are then interested in the
probabilistic study of these matrices. To this end, we randomly
generate unipotent (unitriangular) matrices  and we observe the
number of occurrences of
matrices of substitutions.\\
The construction of unipotent matrices is done as follows:
\begin{enumerate}
    \item Create an identity matrix.
    \item Fill randomly the lower part of the identity
    matrix (strictly under the diagonal) using numbers which follow a
    uniform law in $[0,1]$.
    \item  Multiply those numbers by the range previously chosen.
    \item Test if the  built unipotent matrix satisfies Eq.(\ref{condition de
    substitution}).

\end{enumerate}
We start by giving some examples of our experiment which are
summarized in the table below:

\begin{center}
\begin{tabular}{|c||c||c||c|}
  \hline
  Size & Numbers of drawing & Range of variables & Probability \\
  \hline
  \hline
  \hline
   $[3\times 3]$& $300$  & $[1\cdots 10]$ & $1$  \\
   \cline{3-4}
     &  & $[1\cdots 100]$ & $1$ \\
   \cline{3-4}
   &  &$[1\cdots 10000]$ & $1$ \\   \hline
   \hline
   \hline
   $[4\times 4]$& $275$ &  $[1\cdots 10]$ & $0.0473$  \\
   \cline{3-4}
     &  & $[1\cdots 100]$ & $0.0001$ \\
   \cline{3-4}
   &  &$[1\cdots 10000]$ & $0^+$ \\ \hline
   \hline
   \hline
   $[10\times 10]$& $1500$ &  $[1\cdots 10]$ & $0.0327$ \\
   \cline{3-4}
     &  & $[1\cdots 100]$ & $0^+$ \\
   \cline{3-4}
   &  &$[1\cdots 10000]$ & $0^+$ \\
   \hline

\end{tabular}\end{center}

\vspace{0.5cm}

According to the results obtained, we observe that the
substitutions matrices are not very frequent. However, in meeting
certain conditions such as size, the number of drawings and the
range of the variables, we can obtain positive probabilities that
these matrices appear.\\ Let us note that the smaller the size of
the matrix the more probable one obtains a matrix of
substitution in a reasonable number of drawings.\\
 We also notice that, if
we vary the range of variables, and this in an increasing way and
by keeping unchanged the number of drawings and size, the
probability tends to zero. We also notice that the unipotent
matrices of size 3 are all matrices of approximate substitutions.
This is easy to see because the exponential generating series of
the 3$^{rd}$ column
 will always have the form $\displaystyle{c_k=\frac{x^2}{2!}}$.\\ Thus, we can say
that the test actually starts   from the matrices of  size higher
or equal to 4.

\begin{resultat}
Let $r$ represent the cardinality of the range of variables and
$n\times n$ be the size of the matrix.\\According to the results
obtained; we can say that the probability $p_n$ of appearance of
the matrices of substitutions depends on $r$ and $n$ and we have
the following upper bound:

\begin{eqnarray}
  p_{n} &\leq& \frac{r^{2n-3}}{r^{\frac{n(n-1)}{2}}}
\end{eqnarray}
which shows that

\begin{center}
\begin{eqnarray}
  p_n\longrightarrow 0\quad  &\mbox{as}& \quad n\longrightarrow\infty
\end{eqnarray}
\end{center}

\end{resultat}

\begin{conjecture}One can conjecture that the effect of the
range selection vanishes when n tends to infinity. More precisely:
\begin{eqnarray}
  p_{n} &\sim& \frac{r^{2n-3}}{r^{\frac{n(n-1)}{2}}}
\end{eqnarray}
\end{conjecture}

\end{document}